\documentclass[5p]{elsarticle}
\usepackage[english]{babel}
\usepackage[utf8]{inputenc}
\usepackage{float}
\usepackage{palatino}
\usepackage{amssymb}
\usepackage{pdfpages}
\usepackage{hyperref} 
\usepackage{booktabs}
\usepackage{graphicx}
\usepackage{multirow}
\usepackage{amsmath}
\usepackage{multicol}
\usepackage{txfonts}
\usepackage{natbib}
\begin{document}
\begin{frontmatter}



\title{Introducing {\normalfont\scshape PT-REX },\\the Point-to-point TRend EXtractor}


\author{A. Ignesti\corref{label1,label2}}
\cortext[label1]{INAF-Padova Astronomical Observatory, Vicolo dell’Osservatorio 5, I-35122 Padova, Italy}
\cortext[label2]{Dipartimento di Fisica e Astronomia, Universit\`a di Bologna, via Gobetti 93/2, 40129 Bologna, Italy}

\ead{alessandro.ignesti@inaf.it}

\begin{abstract}
Investigating the spatial correlation between different emissions in an extended astrophysical source can provide crucial insights into their physical connection, hence it can be the key to understand the nature of the system. The point-to-point analysis of surface brightness is a reliable method to do such an analysis. In this work we present PT-REX, a software to carry out these studies between radio and X-ray emission in extended sources. We discuss how to reliably carry out this analysis and its limitation and we introduce the Monte Carlo point-to-point analysis, which allows to extend this approach to poorly-resolved sources. Finally we present and discuss the application of our tool to study the diffuse radio emission in a galaxy cluster. 
\end{abstract}

\begin{highlights}
\item The point-to-point analysis is a reliable method to investigate the physical connection between different components in complex, extended astrophysical sources; 
\item We introduce the Monte Carlo point-to-point analysis which allows to reliably perform the point-to-point analysis also to small or poorly-resolved sources;
\item We present PT-REX, a new software that allows to easily perform point-to-point analysis between radio and X-ray emission. It features several statistical tools to reliably evaluate the spatial correlation for a plethora of scientific cases.
\item As an example, we present here how to use PT-REX to perform the point-to-point analysis on a diffuse radio source in the galaxy cluster RX J1347.5-1145. Due to its architecture, PT-REX can be used to conduct point-to-point analysis on every kind of extended emission, as well as radio and X-ray. 
\end{highlights}


\end{frontmatter}
\section{Introduction}
 It is no wonder that a multi-wavelength analysis can be the best way to investigate an astrophysical system. In the case of extended sources, the study of the spatial correlation between the surface brightness, $I_a$ and $I_b$, produced by two different emission mechanisms, which can be simply expressed as $I_a\propto I_b^k$, can reveal the physical connection between the processes that are taking place in the source. Observing a positive spatial correlation indicates that the two components responsible for the emissions occupy the same volume in the source, whereas the slope $k$ may provide some insights into their physical link. These studies can be carried out by using the point-to-point (ptp) analysis which is, basically, the comparison of surface brightness measured by two different observations made by sampling the extended source with a grid. Under the assumption that each cell of the grid cover the same space of the celestial sphere in each observation, the ptp analysis is more flexible than a comparison between surface brightness radial profiles because it can be seamlessly performed on asymmetrical sources and it can be more responsive to the presence of substructures embed in the extended emission\footnote{The drawback is that, contrary to radial profile which can be easily interpreted by assuming spherical symmetry of the system, understanding the physical meaning of a ptp trend could be not trivial, especially for non-spherical objects.}.
 One of the first applications of the ptp analysis could be found in \citet[][]{Govoni_2001}, where it was featured to study the spatial correlation between radio and X-ray emission in galaxy clusters hosting diffuse radio emission. In this context, the trend between radio and X-ray emission indicates that the thermal plasma is permeated with magnetic field and relativistic particles and it suggests that the radio emission depends on the local properties of the thermal intra-cluster medium (ICM) \citep[e.g.,][for a detailed discussion of this]{Brunetti-Jones_2014}. \\

In this work we present PT-REX, a tool to easily carry out ptp analysis between radio and X-ray emission. We also introduce a new approach, the Monte Carlo ptp analysis, which extends the application of the point-to-point analysis to small, poorly-resolved objects. PT-REX offers the possibility to study the spatial correlation by exploiting a set of different fitting methods to tackle different scientific problems.  This paper is structured as follow. In Section 2 we present the tool, we explain how to use it effectively and we introduce the different fitting methods. In Section 3 we show how to conduct an analysis with PT-REX on a real diffuse raadio source and how the different statistical methods can produce different results. 
\section{{\rmfamily\scshape PT-REX }}
We present the Point-to-point TRend EXtractor ({\rmfamily\scshape PT-REX })\footnote{\nolinkurl{https://github.com/AIgnesti/PT-REX/blob/master/PTREX.tar.xz}}, a flexible Python script to easily carry out the ptp analysis on every kind of extended radio source. PT-REX handles most of the operations with the the Common Astronomy Software Applications (CASA) packages v6.0 \citep[][]{McMullin_2007} developed by the National Radio Astronomy Observatory. We integrated the CASA tools with a variety of Python libraries from Astropy \citep[][]{Astropy_2013, Astropy_2018} and Scipy \citep[][]{Scipy_2020}.
The code is structured in a series of tasks to handle the individual steps of a ptp analysis independently, from defining a grid to sample the radio emission to accurately analyze the data with several statistical methods. A preliminary version of PT-REX has already been used in \citet[][]{Ignesti_2020}, in which we also present the concept of Monte Carlo ptp analysis. Here we present in detail each task and we discuss how to run PT-REX to perform ptp analysis effectively.
\subsection{Data preparation}
PT-REX works by combining radio and X-ray images of an extended source. Therefore, in order to have reliable results, input images must have matching coordinates systems to assure that  the source is mapped by the same sky coordinates. Radio images can be produced with any preferred software, provided that they include information about the beam size and the pixel/arcsec scale in their header to be read with CASA task {\tt imhead}. Concerning the X-ray images, multiple observations can be combined together to improve the count statistic. The X-ray images can be provided as a single exposure-corrected and background-subtracted image in units of surface brightness (e.g., photons cm$^{-2}$ $s^{-1}$) or by providing the count, background and exposure images separately. CASA region files, which are necessary to define the grid and the masks (see Section \ref{samp} and \ref{mask}) can be defined while running PT-REX by using the CASA {\tt imview}. Finally, ancillary information about the calibration error of the radio images and the preferred statistic method (see Section \ref{stat}) have to provided before running the analysis.

\subsection{Sampling algorithm}
\label{samp}
The core of the ptp analysis is the sampling of the diffuse emission. We developed a simple algorithm that converts a rectangular region into a grid that follows the morphology of the radio source. The region is intended to include the source which is going to be sampled, and thus we refer to is as region of interest. The boundaries of the sampling grid, as the cell-size (here intended as the size of the grid cells), the lower threshold in surface brightness to be sampled, and the regions to exclude, have to be provided at the begin of the analysis. In order to have a reliable reconstruction of the radio flux density and to reduce the correlation between contiguous cells, the cell-size has to be large at least as the resolution of the radio image. A larger cell-size can be adopted to increase the signal-to-noise ratio of each cell in the radio and X-ray observations. However, for a given source using larger cells entails a lower number of points (i.e., a lower statistic) to study the spatial correlation that can potentially jeopardize the analysis. A rule of the thumb is that 15 cells, at least, are necessary to sample the diffuse emission to assure a reliable outcome of the analysis, so a compromise between resolution of the grid and signal-to-noise has to be found. Finally, the threshold defines the lowest value of surface brightness that is going to be sampled by the grid and it is expressed in unit of Jy beam$^{-1}$.\\
\begin{figure}[t!]
 \includegraphics[width=.45\textwidth]{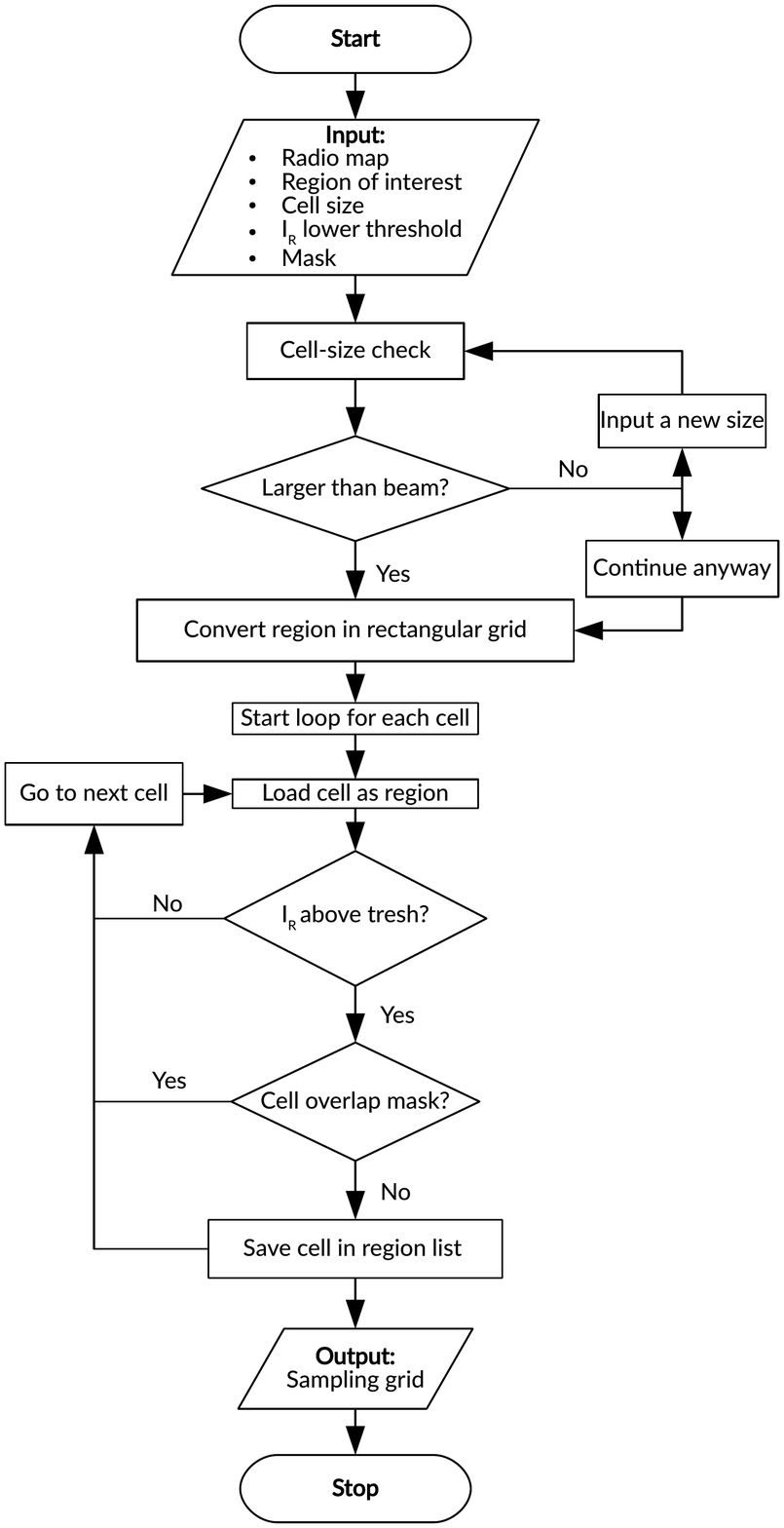}
 \caption{\label{grid}Flowchart of sampling algorithm.}
\end{figure}
The sampling algorithm is quite straightforward and it described in Figure \ref{grid}. After a preliminary check on the cell-size, the region of interest is converted in a rectangular grid. At this step, the coordinates of each cell are defined in the pixel units of the radio image. Then the radio surface brightness and the position with respect to the mask are evaluated for each cell of the grid, starting from the bottom-left to the top-right. These checks are done with the CASA task {\tt imstat}. All those cells that do not meet the requirements (i.e., those measuring a radio surface brightness below the threshold or overlapping with the masked regions) are excluded, whereas the others are converted in J2000 coordinates and finally stored in a region file, which is the final output of the routine. Every sampling grid can be displayed on the top of the radio image by using the CASA {\tt viewer} and it can be further modified manually to better adapt to the science case. After having defined a sampling grid, we can use it to compare the radio and X-ray emissions. 

\subsection{Single-mesh analysis}
\begin{figure}[t!]
 \includegraphics[width=.45\textwidth]{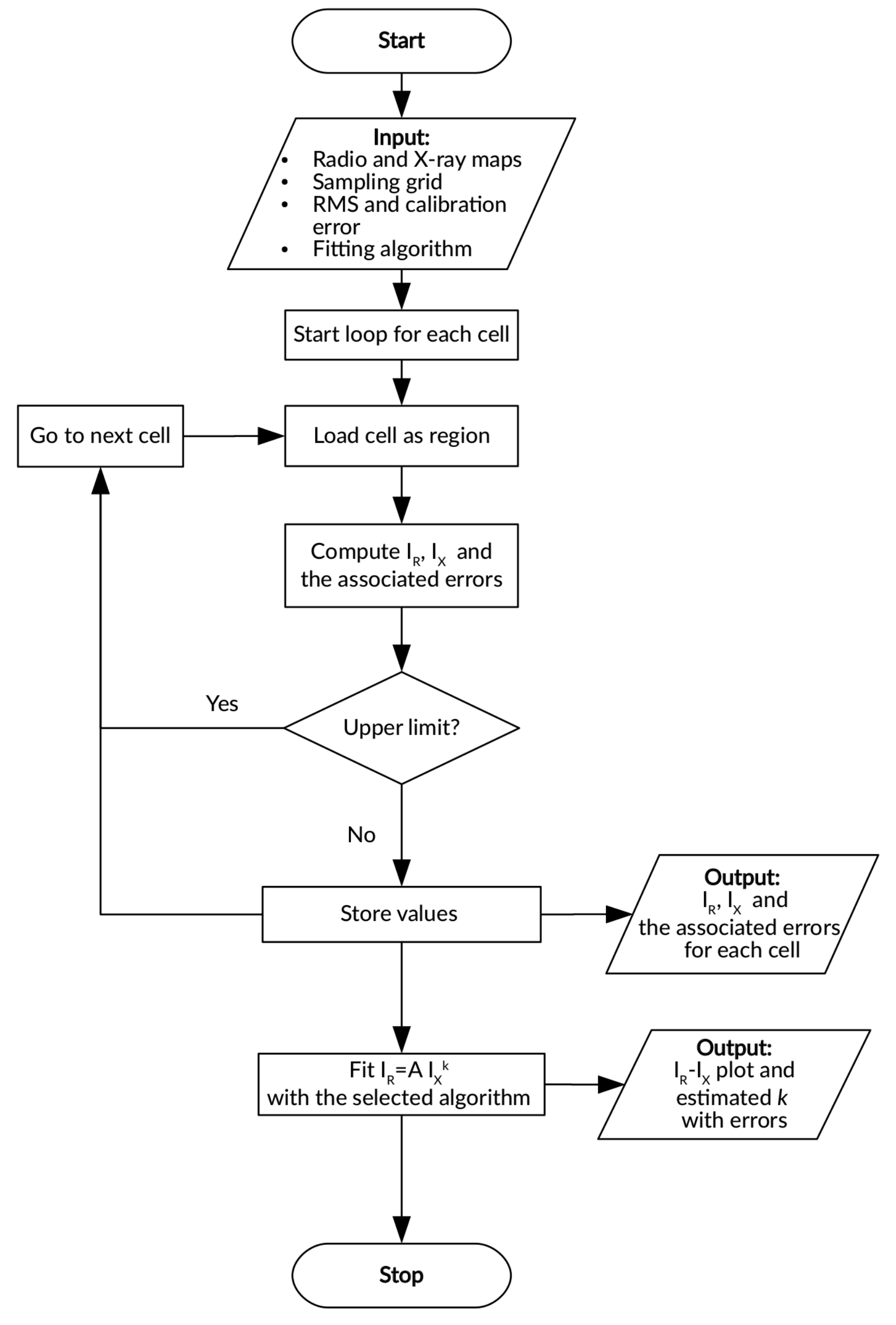}
 \caption{\label{SMptp}Flowchart of the SMptp routine.}
\end{figure}

We define a single-mesh ptp (SMptp) analysis a ptp analysis carried out by using only a mesh to sample the radio emission. It is composed of two steps: first the radio and X-ray surface brightness are measured in each cell of the grid, then they are compared as $I_\text{R}$ vs $I_\text{X}$  to evaluate the spatial correlation. We present the flowchart of the process in Figure \ref{SMptp}. The routine that collect the values of $I_\text{R}$ and $I_\text{X}$ is straightforward. For a given sampling grid, which was previously created by using the sampling algorithm, $I_\text{R}$ and $I_\text{X}$ are evaluated for each of its cells. Because the cells are defined in J2000 coordinates according to the radio map coordinate system (see Section \ref{samp}), by using maps with matching coordinates each cell will sample the same direction of the sky for both the radio and X-ray images. However, we note that if the X-ray map pixel-size is larger than the radio map one, then there could be a mismatch between the position of the grid cells in the two images due to the different sampling of the sky coordinates. In this case, we suggest to check how the sampling grid is positioned on the X-ray map before running the SMptp analysis and, if possible, to regrid the X-ray map to match the size and the pixel-size of the radio one.\\

For the radio image, the flux density in each cell is measured with the CASA task {\tt imstat} and the converted in $I_\text{R}$ by dividing for the area of the cell, $\Omega_\text{c}$, in units of arcsec$^2$. The associated error, $\sigma_\text{R}$, is computed as;
\begin{equation}
 \sigma_\text{R}=\frac{\sqrt{\left(f\cdot S \right)^2+\left(\text{RMS}\cdot\sqrt{\Omega_\text{c}/\Omega_{B}} \right)^2}}{\Omega_\text{c}}
 \label{err}
\end{equation}
where $f$ and RMS are, namely, the amplitude calibration error and the root mean square of the radio image provided by the user, $S$ is the flux density measured in the cell and $\Omega_{B}$ is the beam area. Since every cell has the same size, the second term of Equation \ref{err} is the same for all and it becomes significant only for those cells with a low radio surface brightness. As for the X-ray images, when several Chandra observations of the same cluster are involved, we compute the total $I_\text{X}$ of a cell as
 \begin{equation}
  I_\text{X}=\frac{\sum N_\text{cnt,i} - \sum N_\text{bkg,i} }{\sum q_\text{exp,i}}\frac{1}{\Omega_\text{c}}=\frac{\sum \left(S_\text{X,i}\cdot q_\text{exp,i}\right)}{\sum q_\text{exp,i}}\frac{1}{\Omega_\text{c}}
  \label{Itot.math}
 ,\end{equation}
where
 \begin{equation}
  S_\text{X,i}=\frac{N_\text{cnt,i}-N_\text{bkg,i}}{q_\text{exp,i}}
 \end{equation}
 is the flux measured for the $i$-th Chandra observation, $\Omega_\text{c}$ is the angular area of the cell in units of arcsec$^2$
, and $N_\text{cnt,i}$, $N_\text{bkg,i}$ (in units of counts), and $q_\text{exp,i}$ (in units of counts cm$^{2}$ s photons$^{-1}$) are, respectively, the values measured on the counts, the background, and the exposure map of the $i$-th Chandra observation. When no background or exposure maps are provided, their values are set to, respectively, 0 and 1 for each cell\footnote{This entails that, by providing only the count image, this procedure is valid for every other kind of observation where the signal can be divided into counts, e.g. an optical image. This makes PT-REX virtually able to compare radio emission with other kinds of emissions, as well as X-ray emission.}. We derive the associated errors on $S_\text{X,i}$ by assuming a Poisson error for $N_\text{cnt,i}$ and $N_\text{bkg,i}$ and computing the error propagation of Equation \ref{Itot.math}. During this phase, the sampling is further refined by excluding all these cells that measure negative values of $I_\text{R}$ and $I_\text{X}$ or upper limits in radio and X-ray, i.e. those values with relative uncertainties greater than 100$\%$. As a caveat, we note a limit of the current sampling algorithm. Due to the fact that the sampling is mainly driven by the signal-to-noise ratio of the radio image, the final grid may include cells with an X-ray emission close to the noise level which may negatively impact on the fit. In this case, we suggest to either increase the cell-size to reach a compromise between the X-ray signal-to-noise and the resulting sampling cells number, or to manually remove the most critical cells by using the CASA task {\tt imview} to display and modify the grid.\\

Once the values of $I_\text{R}$ and $I_\text{X}$ have been calculated, the fitting can begin. We fit the $I_\text{R}$-$I_\text{X}$ distribution with a power-law relation as:
\begin{equation}
 I_\text{R}=A\cdot I_\text{X}^k
\end{equation}
We propose a set of different fitting algorithms to measure $k$ and its associated error $\sigma_{k}$ (see Section \ref{stat}). We also provide a direct estimate of the linear correlation in the logarithmic space by estimating the Spearman and Pearson ranks with the {\tt scipy.stats} library. At the end of this routine a data file with $I_\text{R}$ and $I_\text{X}$ and relative errors is produced as output. These values can be used for further analysis, e.g. to be examined with a fitting method which is not included in PT-REX \citep[for example LIRA, ][]{Sereno_2016} or, by combining multiple SMptp analysis of the same object observed at different radio frequencies, to study the spatial correlation between spectral index and $I_\text{X}$. In addition to the data file, PT-REX  produces a simple plot with the data and the best-fit line with the interval which has the $95\%$ chance of containing the true regression line. \\

As a caveat, we note that the reliability of the SMptp analysis is limited by the assumption that any sampling grid provides an unambiguous reconstruction of the real surface brightness of the source. Such an assumption is valid for those sources that can be sampled by large number of cells, i.e. where radio emission is well resolved by the observation, but it may not be true for the other smaller or poorly-resolved objects. In the next section we discuss how the ptp analysis can be extended also to these cases.

\subsection{Monte-Carlo analysis}
Extended sources may contain smaller substructures (e.g. bright filaments embed in the extended emission or surface brightness gradients). If the resolution of the observation is sufficient to fully resolve them, i.e. the angular resolution is lower than half of the angular scale of these features, the result of the SMptp analysis will not depend on the sampling grid because the surface brightness will be reliably reconstructed by every possible combination of cells. Otherwise, if these substructures can not be properly sampled by the observation/grid, the resulting SMptp analysis will be fatally biased by the choice of the sampling grid. In this latter case, an approach more complex than the SMptp is required to estimate the trend of the spatial correlation. \\

\begin{figure}[t!]
 \includegraphics[width=.5\textwidth]{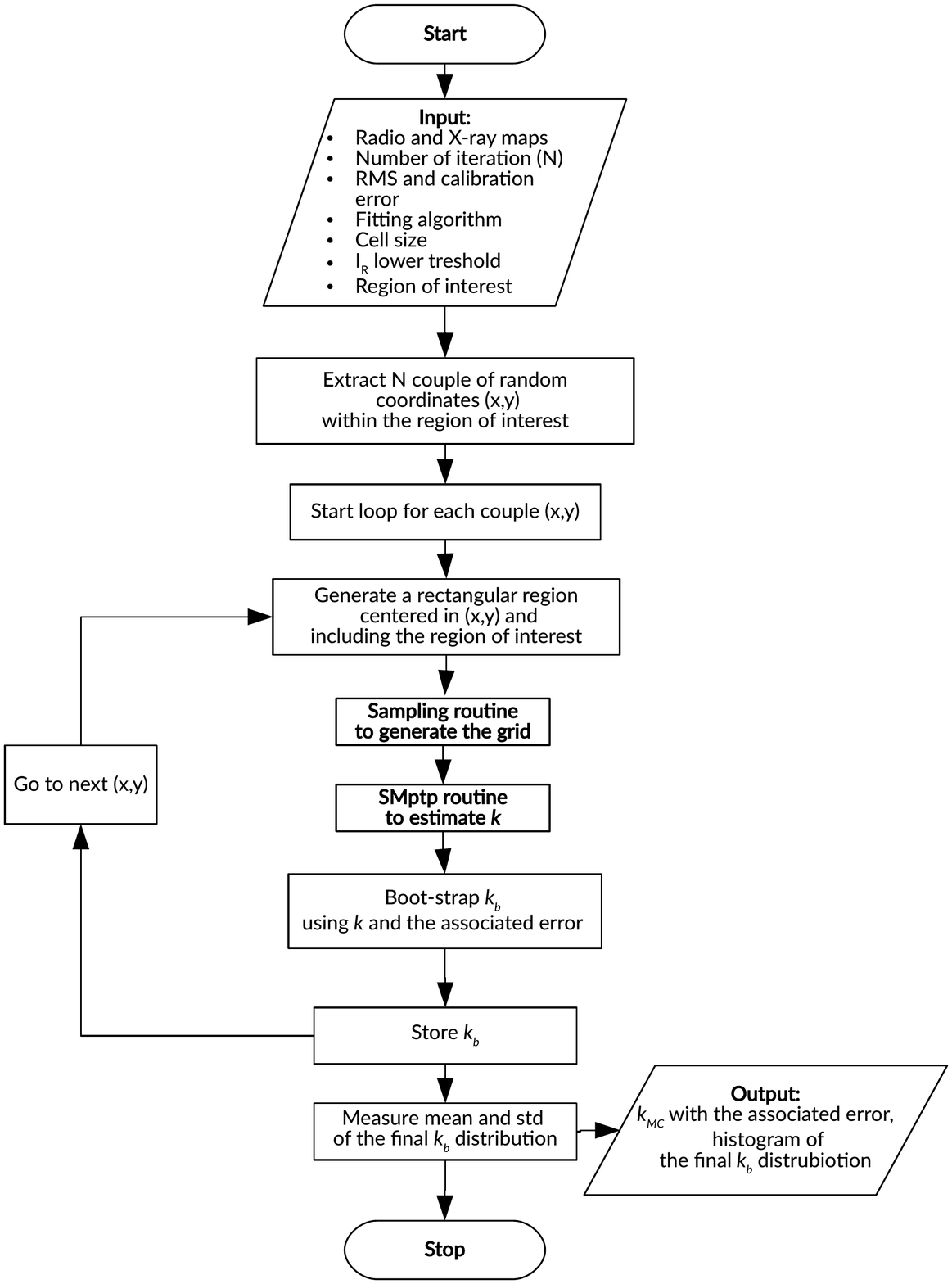}
 \caption{\label{MCptp}Flowchart of the MCptp routine.}
\end{figure}

The major feature introduced by {\rmfamily\scshape PT-REX } is the possibility to use an automatic, randomly-generated sampling routine to combine several SMptp analysis into a Monte Carlo ptp (MCptp) analysis. By repeating several cycles of SMptp analysis with randomly-generated grids, PT-REX produces a distribution of values of $k$ that describe its parameter space, thus it allows us to reliably estimate the trend (and its uncertainties). This routine makes uses of the sampling algorithm and the SMptp routine presented in the previous sections. We present the flowchart of the process in Figure \ref{MCptp}. After setting the number of Monte Carlo iterations ($N$), the region of interest, the cell-size and the $I_\text{R}$ lower threshold are set, the Python function {\tt numpy.random} is used to generate a number $N$ of coordinates $(x,y)$ within the region of interest. These coordinates are used to define $N$ different, rectangular regions centered in $(x,y)$ and large enough to include the original region of interest, hence the radio source. Then these $N$ regions are converted in sampling grids by the sampling algorithm and used to carry out the corresponding SMptp analysis to measure $k\pm\sigma_k$. \\ 
At the end of each cycle, a random value $k_b$ is extracted from a normal distribution centered in $k$ with sigma $\sigma_k$ to be finally stored. This boot-strapping procedure enables us to transpose the error of each individual fit in the final $k_b$ distribution. After all the $N$ different grids have been exploited, the result of the MCptp analysis, $k$, is computed as: 
\begin{equation}
k=\overline{k}_\text{b}\pm\sigma_{k_\text{b}} 
\end{equation}
where $\overline{k}_\text{b}$ and $\sigma_{k_\text{b}}$ are the mean and the standard deviation of the distribution of boot-strapped $k_b$ obtained at the end of each cycle. Usually SMptp and MCptp performed with the same parameters (mask, starting region, cell-size and lower threshold) should be consistent within 1 $\sigma$. If not, this could be due to 1) a very peculiar choice of the SMptp mesh or 2) the presence of unmasked field source that are erroneously, yet consistently, included in the sampling. An histogram of the $k_b$ distribution is produced in output, which may point out additional information about the source. On the one hand, observing a dispersion significantly larger than the SMptp uncertainties indicates that the random sampling affected the estimate, which is sign of a poor sampling of the radio emission. On the other hand, an asymmetrical distribution may indicate the presence of a secondary component in the radio source that, for fortuitous combinations of cells have pivoted the fit. For instance, low-brightness components or strong X-ray point-sources embed in the radio emission can produce a negative skewness in the distribution, whereas the presence of point sources with a strong radio and X-ray emission can induce a positive skewness.\\

MCptp analysis is advised for those sources that can be sampled with $<30$ cells. We suggest setting the number $N$ to a minimum of 100 to adequately probe the parameter space of $k$. We note that the number of iterations, the sizes of the region of interest and of the cells and the number of X-ray maps involved in the analysis affect severely the duration of the procedure. As a basic rule, mind that adjusting the cell-size by a factor $f$ changes the number of sampling cells of $\sim1/f^2$, which, in turns, modifies the processing time of the same amount. PT-REX does not feature parallel processing, thus it performance in terms of processing time may vary depending on the available hardware. A scientific application of MCptp analysis to study radio mini-halos in galaxy clusters is presented in \citet[][]{Ignesti_2020}.    
\subsection{Generate a mask}
\label{mask}
Field sources and those embed in the radio emission not associated with the extended source (e.g. central radio galaxies) can jeopardize the results of the ptp analysis. When the subtraction of those sources from the data is not possible, they can be masked and excluded from the ptp analysis. \textsc{PT-REX } includes a tool that allows the user to produce masks for the analysis simply by providing the regions intended to be excluded. The regions are used to define a matrix with the same size of the radio image that allows the sampling routine to recognize and exclude any unwanted sources. Typically, there are two kind of sources that have to be masked, 1) those embed in the extended emission, as physically located within the source but its emission is not related it, or background and foreground sources along the line of sight or 2) those that are outside the diffuse emission but close to the region of interest. As a simple rule of the thumb, we suggest to define carefully the regions to be masked within the source and the region of interest. On the on hand, define a mask that exceeds the size of the unwanted source can lead the sampling routine to exclude a number of cells larger than the necessary, and thus to reduce the number of point to evaluate the spatial correlation. On the other hand, a mask smaller than the source should not contain the contribution of the source in the extracted surface brightness, thus jeopardizing the analysis. As for the sources close to the region of interest, they can be problematic during a MCptp analysis. At the begin of the cycle a new region of interest is defined and some of them can be erroneously included and sampled. So we suggest to adopt large masks to safely account for their presence within $2\times$, both in width and in height, the size of the region of interest. 
\subsection{Fitting algorithms}
\label{stat}
 Fitting $I_\text{R}\propto I_\text{X}^{k}$ is a crucial part of the ptp analysis, and different scientific problems may require different statistic methods to evaluate the trend. For this reason, {\rmfamily\scshape PT-REX } includes a range of fitting algorithms: 
\begin{itemize}
 \item Least squares (LS): data can be fitted with a power-law relation $I_\text{R}=A\cdot I_\text{X}^k$ by using the least-squares method with the {\tt scipy.optimize.curve$\_$fit} method. Only the uncertainties on $I_\text{R}$ are taken into account. This method estimates the best-fitting parameter of the power-law which minimize the distance from the data, under the assumption that the data intrinsically follow a power-law distribution and the scatter is only due to observational errors. Due to its assumptions, this method can be biased by outliers that can pivot the fit. For a physical point of view, the assumption of an intrinsic, perfect correlation between the two quantities may be questionable. In a complex, physical system the apparent correlation between two quantities can depend on a third, unknown factor. In this case, an internal scatter of the data is expected regardless of the quality of the observations and, thus, the base assumptions of this method can lead to biases in the scientific conclusions. Therefore, although we include this method for the sake of completeness because it has been largely used in literature, we advise to use it cautiously;
 \item  BCES orthogonal and bisector: The fitting method is the bivariate correlated errors and intrinsic scatter (BCES) presented in \citet[][]{Akritas_1996}. The BCES regression offers several advantages compared to ordinary least squares fitting, as measurement errors on both variables and it can account for an intrinsic internal scatter of the data. The fitting is performed by the {\tt bces} module\footnote{{\tt https://github.com/rsnemmen/BCES}}. $I_\text{R}$ and $I_\text{X}$ are transposed in the logarithmic space before being fitted as $\text{log}I_\text{R}=k\text{log}I_\text{R}+\text{log}A$. By default, we assume that the errors on $I_\text{R}$ and $I_\text{X}$ are not correlated. We included in PT-REX both the orthogonal method and the bisector method, which compute the symmetric lines constructed from the BCES best-fits of $(I_\text{R}|I_\text{X})$ and $(I_\text{X}|I_\text{R})$;
 \item LinMix: this is a Bayesian method to account for measurement errors in linear regression introduced in \citet[][]{Kelly_2007}. This method allows for heteroscedastic and possibly correlated measurement errors and intrinsic scatter in the regression relationship. The method is based on deriving a likelihood function for the measured data, especially for the case when the intrinsic distribution of the independent variables can be approximated using a mixture of Gaussian functions. LinMix incorporates multiple independent variables, nondetections, and selection effects (e.g., Malmquist bias). We implemented this algorithm with the {\tt linmix} module\footnote{{\tt https://github.com/jmeyers314/linmix}}. This method derives a likelihood function for the data, thus the best-fit slope is estimated from the mean of the posterior distribution. To run this method, a number of chains for the bayesian algorithm, {\tt n$\_$chain}, and the number of gaussian to build the prior, {\tt K}, has to be defined by the user. This method is significantly more time-consuming than the other options. We advise to use it when a large number of cells is involved and the chose of the lower threshold is expected to impact on the fit. 
\end{itemize}
A detailed discussion about the best fitting strategy to adopt for different science cases can be found in \citet[][]{Isobe_1990, Akritas_1996, Kelly_2007, Hogg_2010}.
\section{Application to a scientific case}
We present here the application of {\rmfamily\scshape PT-REX} to study the correlation between radio and X-ray emission in the mini-halo in the RX J1347.5-1145 galaxy cluster. Radio mini-halos are diffuse radio sources observed at the center of relaxed clusters and whose origins are still debated. Although they strongly differ, all the models which have been proposed to explain the origin of the relativistic, radio-emitting electrons agree on linking the properties of the non-thermal particles to the thermal gas \citep[e.g.,][for reviews]{Brunetti-Jones_2014,vanWeeren_2019}. Therefore, studying the spatial correlation between radio and X-ray emission is crucial to investigate this physical connection and to discriminate between the different scenarios. The mini-halo in RX J1347.5-1145 is elongated, where the brightest part located toward north with a fainter extension toward south. The diffuse radio emission surrounds the bright radio galaxy GALEX J134730.-114509, and two large extended radio sources are placed side by side with it. The X-ray emission exhibits an elliptical shape, although more symmetric than the radio emission, with the major axis aligned along northwest-southeast. Previous X-ray studies unveiled the complex dynamical status of this cluster where shocks and merging sub-clusters coexist with a sloshing cool-core \citep[e.g.,][]{Mason_2010,Gitti_2007c,Johnson_2012,Kreisch_2016}. Therefore, due to the complex morphology of this system, the ptp analysis is the best suited to study the connection between radio and X-ray emission. 

\begin{figure}[h!]
\centering
 \includegraphics[width=1\linewidth]{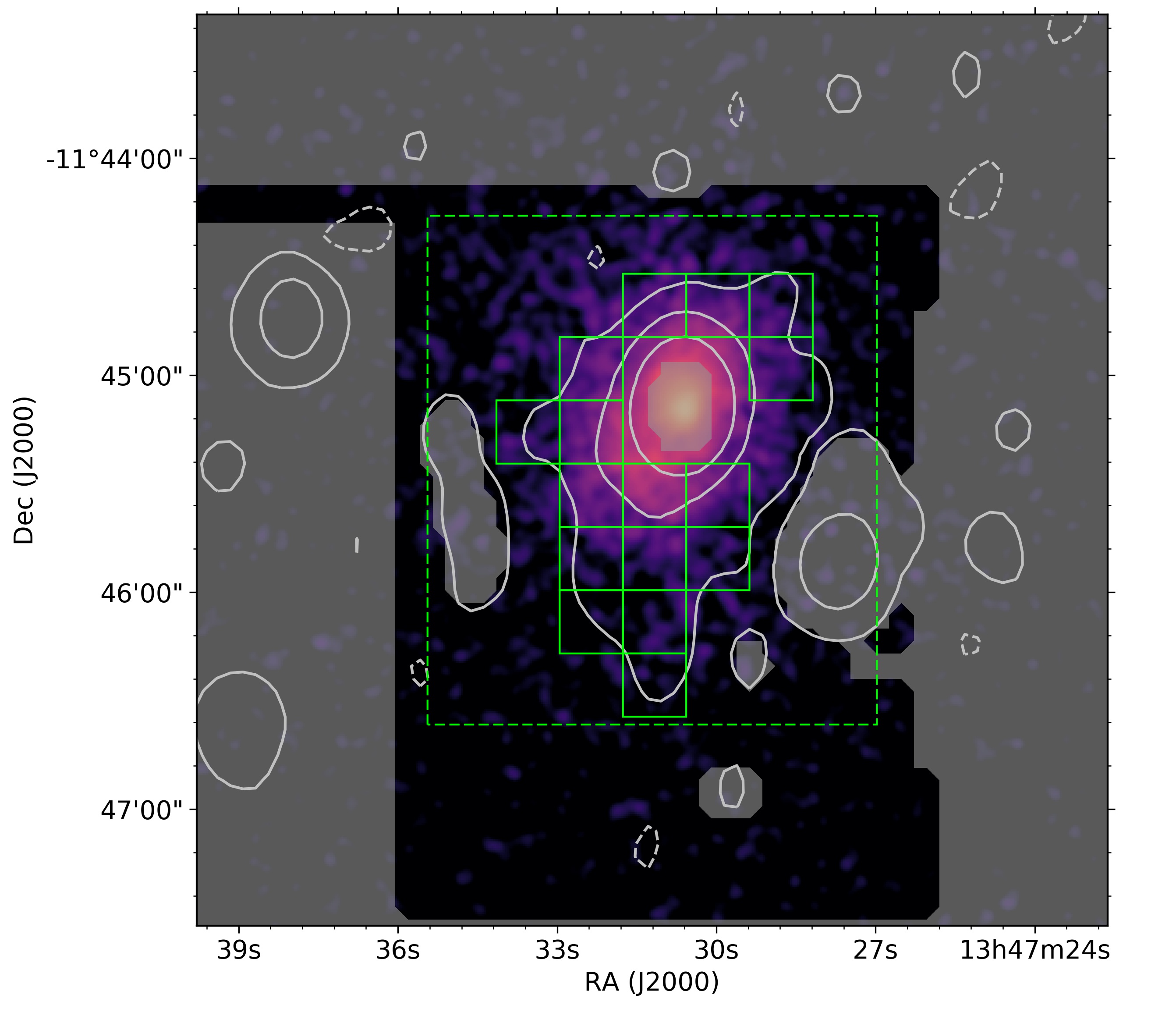}
 \caption{\label{map}X-ray image of RXJ1347.5-1145 in the 0.5-2.0 keV band with the -3, 3, 24, 96$\sigma$ level countours of the VLA radio image from \citet[][]{Gitti_2007b} (1$\sigma$= 0.04 mJy beam$^{-1}$). We report here the region of interest (green dashed), a random smpling grid (green continuous) and the mask (grey).} 
\end{figure}

\begin{figure*}[t!]
\raggedcolumns
 \begin{multicols}{2}
\includegraphics[width=1\linewidth]{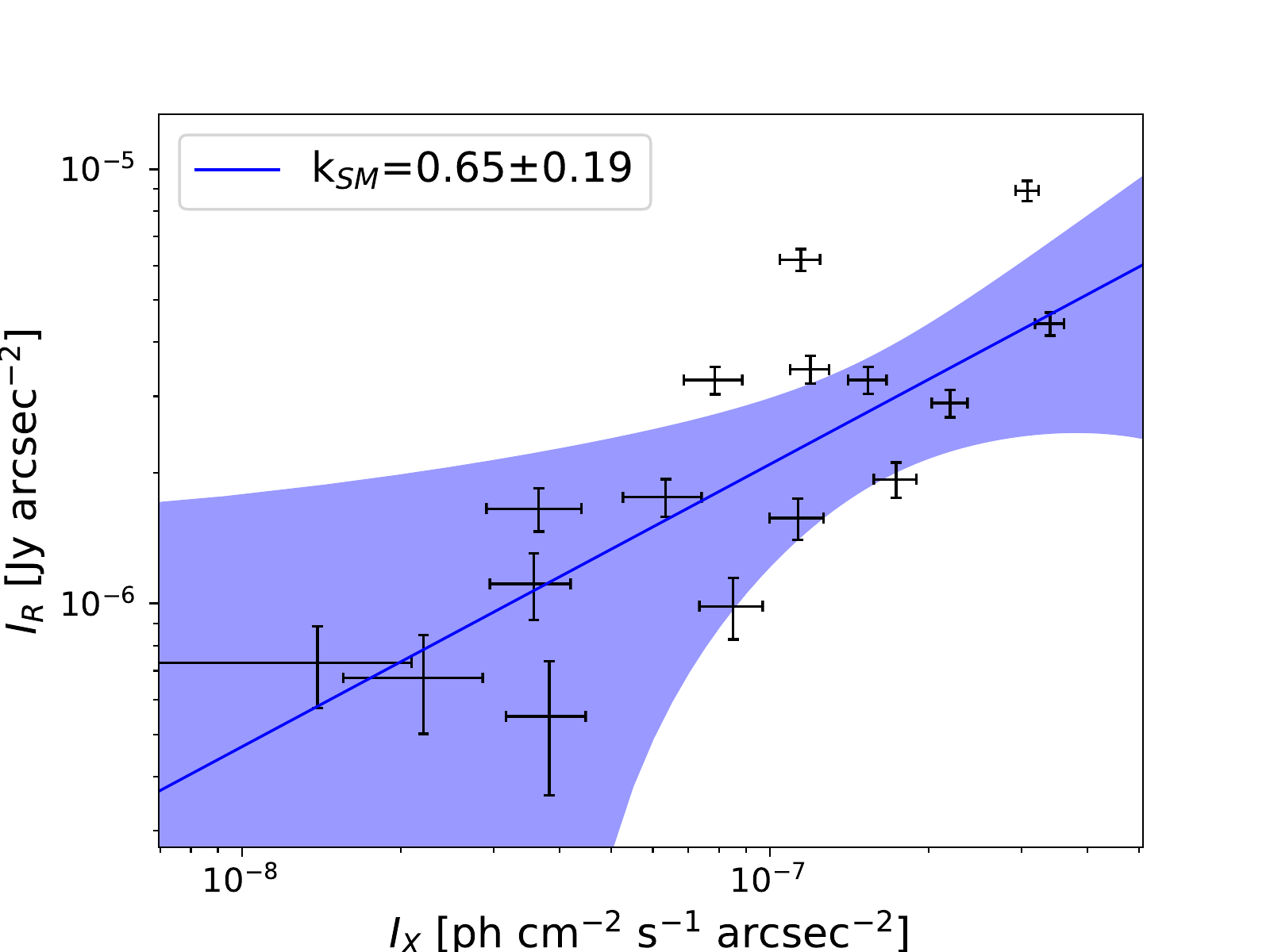}
\includegraphics[width=1\linewidth]{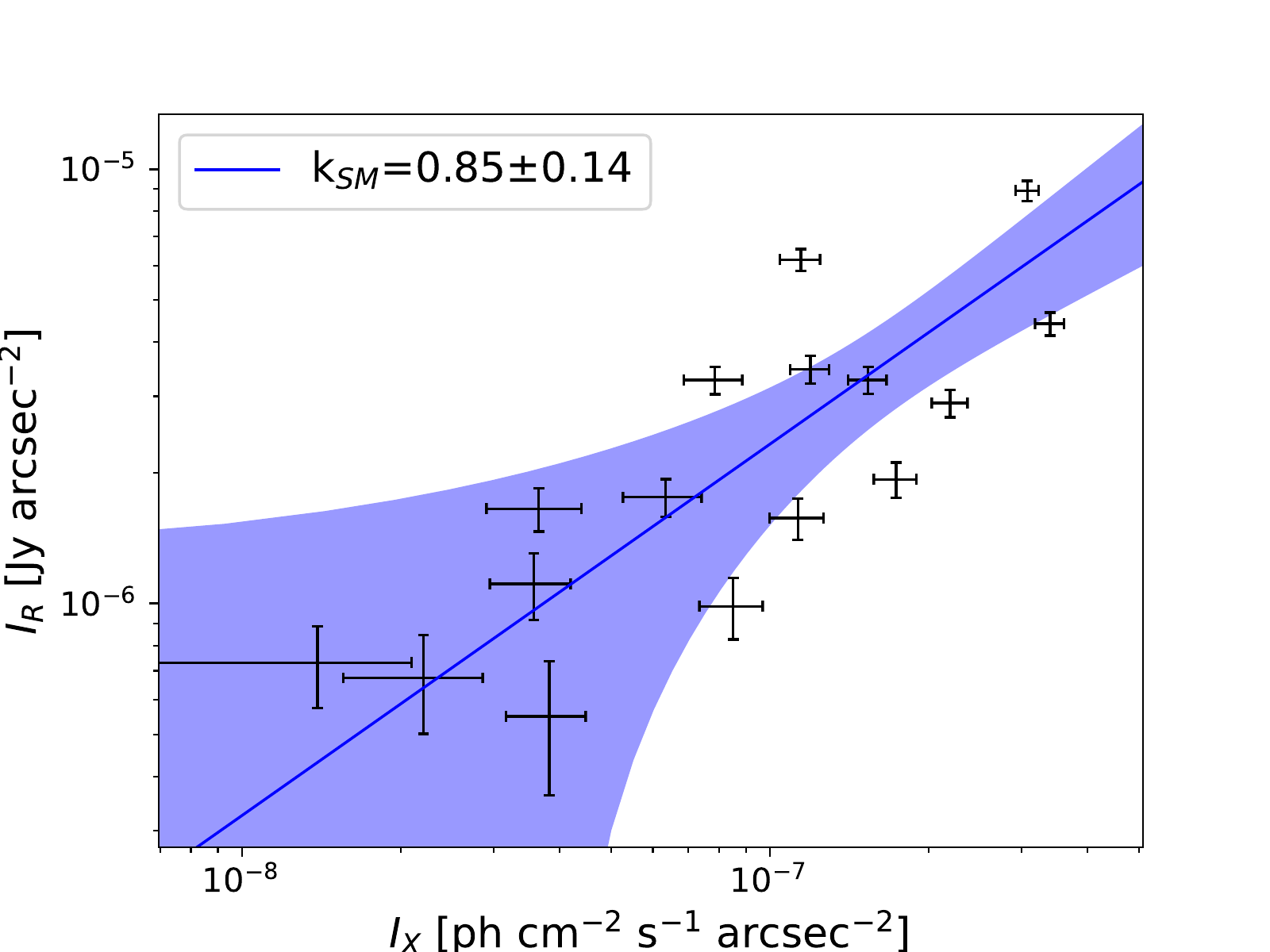}
 \end{multicols}
 \raggedcolumns
 \begin{multicols}{2}
\includegraphics[width=1\linewidth]{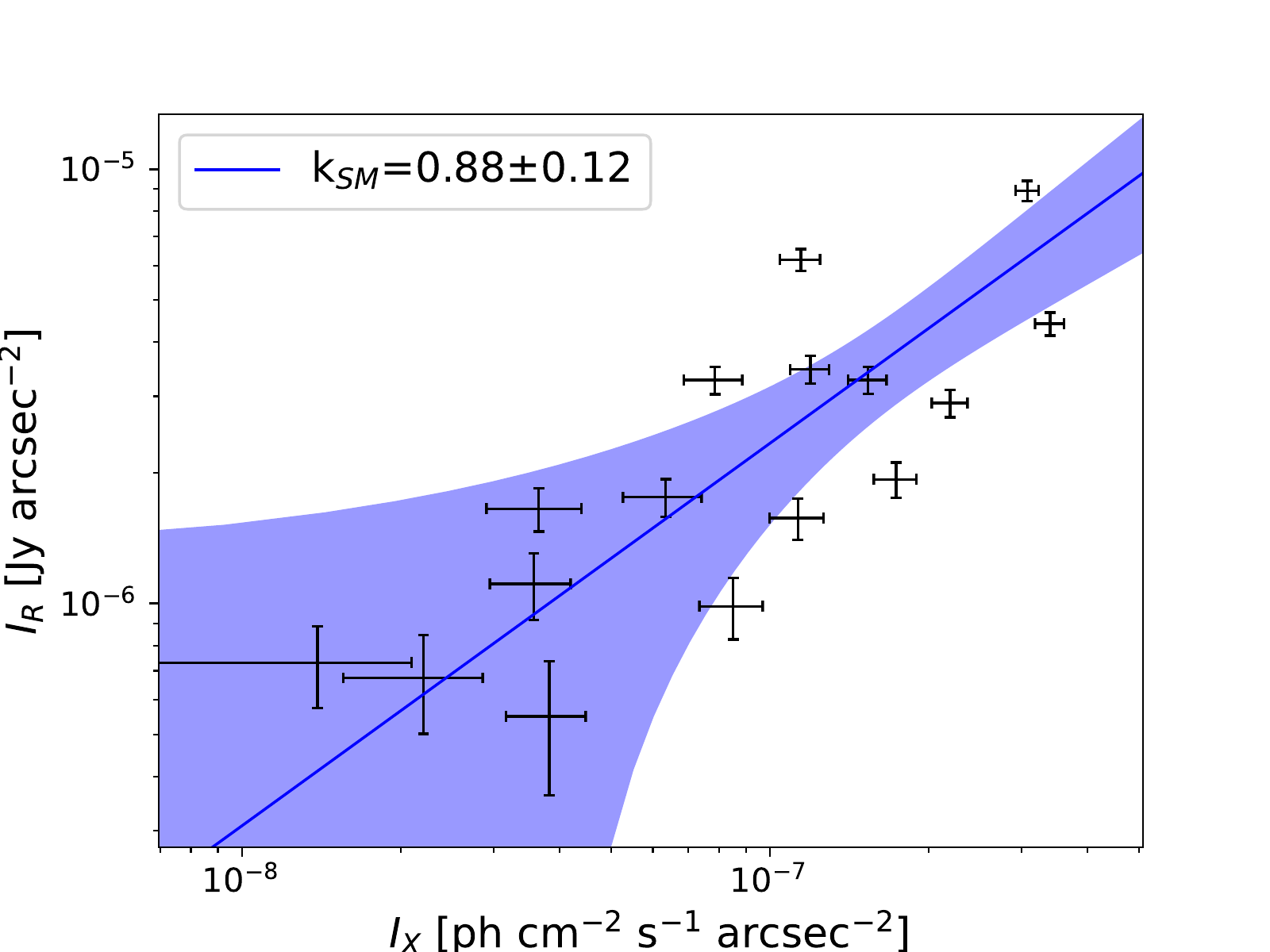}
\includegraphics[width=1\linewidth]{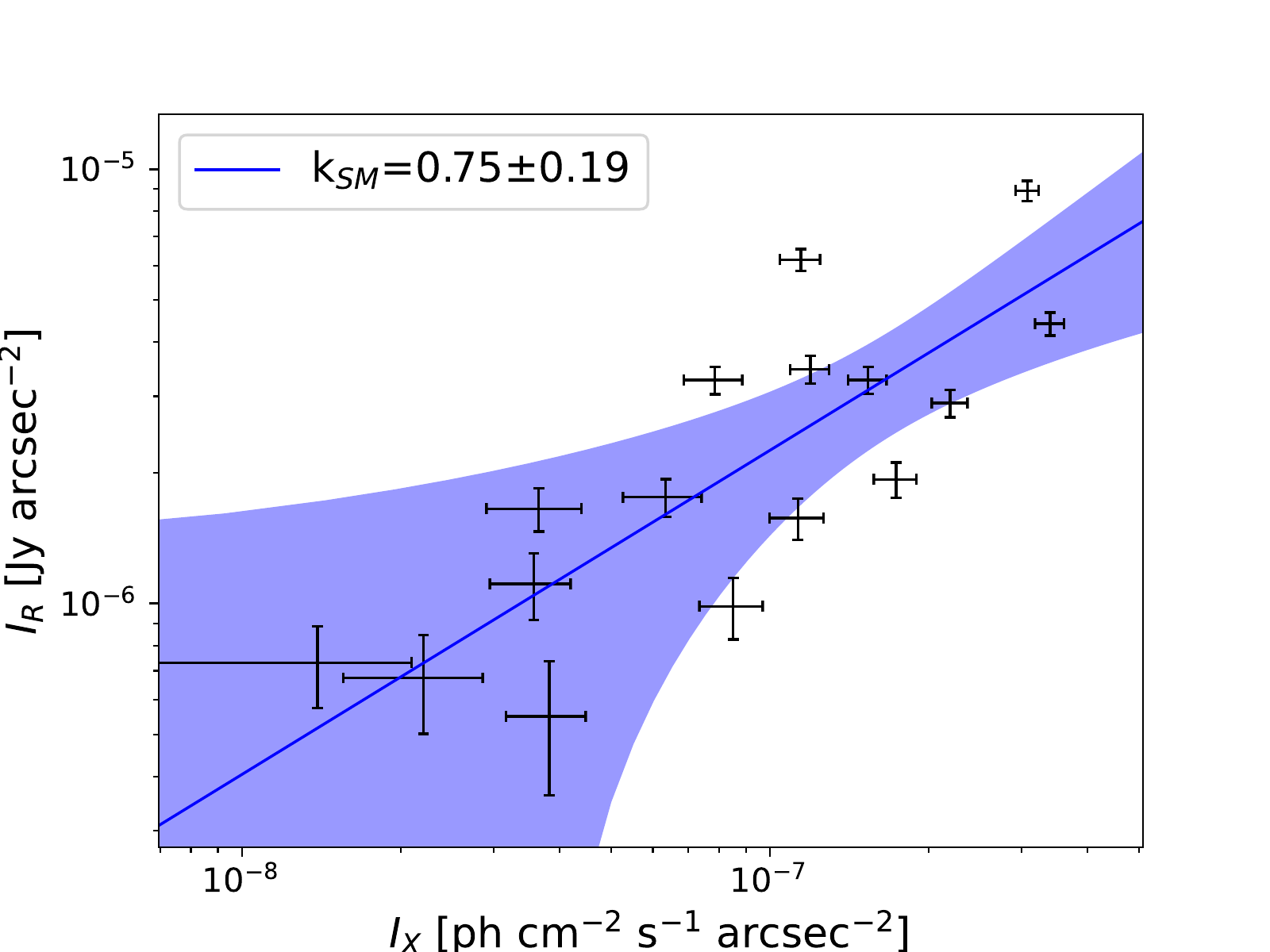}
 \end{multicols}
 \caption{\label{sm} Results of the SMptp analysis carried out with the grid shown in Figure \ref{map} with the different fitting algorithm: Least squares (top-left), BCES orthogonal (top-right), BCES bisector (bottom-left), LinMix (bottom-right). We report the 95$\%$ confidence interval.}
\end{figure*}

\begin{figure}[t!]
\centering
 \includegraphics[width=1\linewidth]{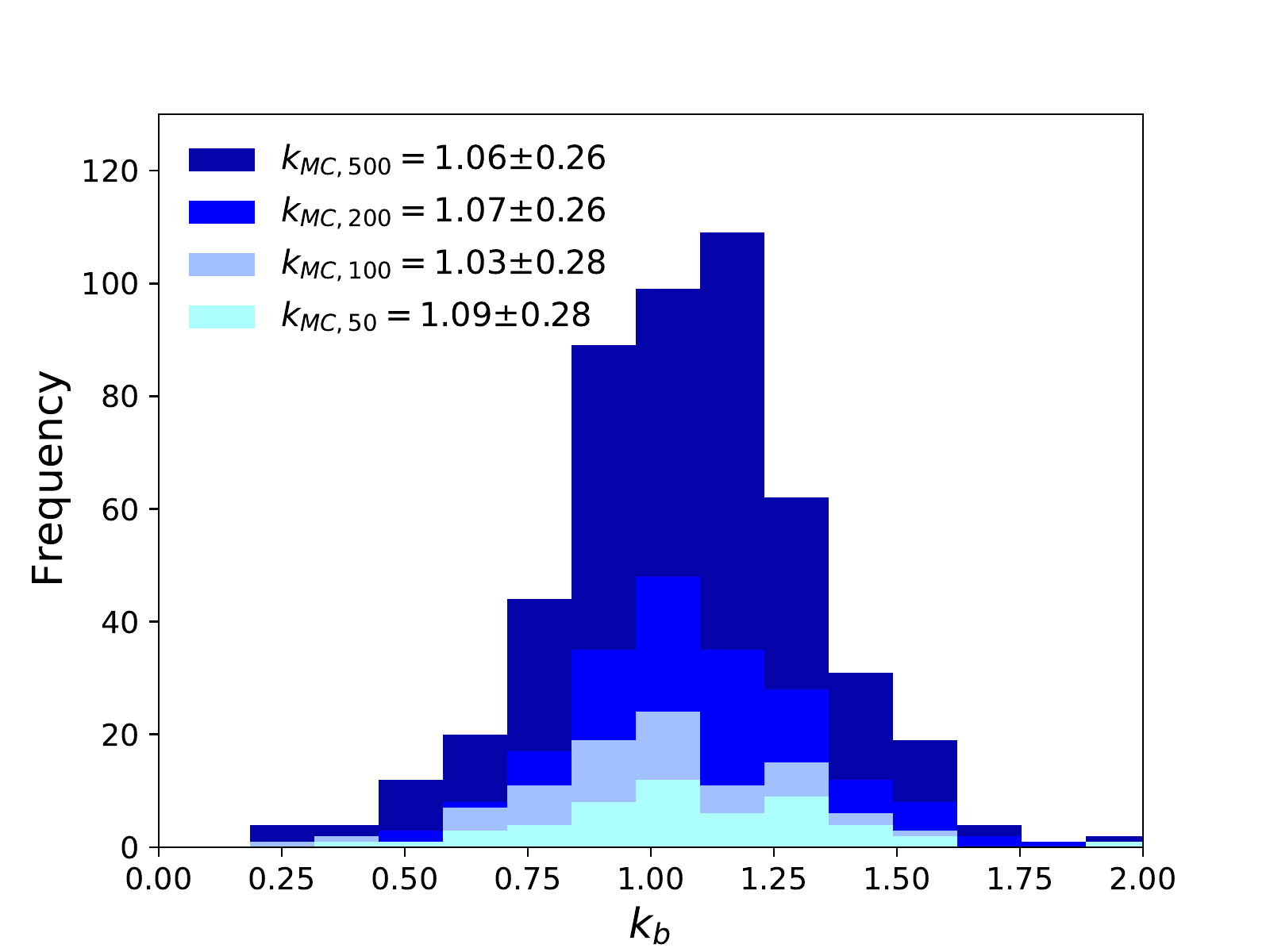}
 \caption{\label{mc} Histogram with the distribution of $k$ produced by 500 iterations of the MCptp routine. We individually report the mean and standard deviation of the first 50, 100, 200 iterations and of the total distribution.}
\end{figure}

We combined the Very Large Array observation at 1.4 GHz presented in \citet[][]{Gitti_2007b} (RMS=0.04 mJy beam$^{-1}$, beam 17.7$''\times13.6''$) with an X-ray image in the 0.5-2.0 keV band produced from the archival Chandra observation 2222 (PI Khan, exposure time 100 ks), which we processed following the standard analysis procedure\footnote{{\tt https://cxc.cfa.harvard.edu/ciao/threads/index.html}}. We used the 3$\times$RMS level contours of the VLA image to define the region of interest, which indicates to PT-REX the size and the position of the radio source, and the mask (Figure \ref{map}). We carefully defined the mask within the region of interest by making sure to remove the emission of the central galaxy and of the field sources, whereas we adopted a more crude approach adopted outside the region by masking the other sources with large squares. Then we define a first grid to sample the emission above the 3$\sigma=0.12$ mJy beam$^{-1}$ level of the radio image. The beam area is 273.6 arcsec$^2$ ( 23 pixel), so we use cell with a 17.5$''$ size and a total area of 289 arcsec$^{2}$ (25 pixel). The resulting grid shown in Figure \ref{map} is composed of 16 cells which were then used in a SMptp analysis. The Spearman and Pearson ranks that we measured are 0.79 and 0.8, respectively, which indicate a strong linear correlation between $I_\text{R}$ and $I_\text{X}$. This indicates that the ICM component responsible of the radio and X-ray emission occupy the same volume, which is in agreement to what is observed in radio mini-halos. In Figure \ref{sm} we report the resulting SMptp analysis carried out with the algorithms implemented in PT-REX. Albeit the large errors, they are all consistent within 1 $\sigma$ and indicate a sub-linear correlation between radio and X-ray emission. 
However, the low number of cells indicates that the source is poorly resolved, so our results may be biased by the grid which we used. Therefore we performed also a MCptp analysis to test this possibility and to better constrain $k$. We present here the results of 500 iterations of MCptp analysis carried out with the BCES orthogonal fit. We used the mask and region of interest presented in Figure \ref{map} with the same cell size and threshold adopted for the SMptp analysis. In Figure \ref{mc} we report the mean and the standard deviation of the final distribution and of the first 50, 100 and 200 iterations.

In Table \ref{res} we report the results of the all the ptp analysis carried out. All the method produced estimates of $k$ that are in agreement within 1$\sigma$, although the MCptp estimates a linear trend. All of the SMptp analysis resulted in a sub-linear trend and, amongst them, the LS fit estimated the flattest trend because it was pivoted by the two points at the extremes of the distribution (Figure \ref{sm}, top-left panel). The two BCES estimators produced similar results, whereas the the LinMix fit (which was performed with 100 chains and a prior composed of two gaussians) returned a slightly flatter value of $k$, albeit with larger errors. By comparing the SMptp analysis performed with the BCES orthogonal algorithm with the MCptp, we can conclude that the grid we used in the former was slightly biased toward flatter values, most likely by the large errors of the low-brightness points. The uncertainties of the MCptp are larger than the ones of the SMptp, regardless of the fitting method. This indicate that the random sampling has significantly affected the estimate of $k$, i.e. it confirms that the observation does not fully resolve the mini-halo and there may be sub-structures embed in the emission likely associated with the features (such as shocks, sloshing and merging sub-clumps) of the thermal ICM. Therefore, by featuring PT-REX we could comprehensively study the spatial correlation between diffuse radio and X-ray emission in RXJ1347.5-1145, finding that they are linked by a linear trend. This implies that the distribution of magnetic field and relativistic particles is deeply bounded to the distribution of the thermal plasma and, therefore, it may suggest that the diffusion and re-acceleration of the radio-emitting electrons depends mostly on the local properties of the thermal gas (e.g. its turbulence). Our result also suggests that the current observations may be missing some components of this radio source, which are instead glimpsed by the random sampling. The scientific interpretation of this result in the context of origin of the radio emission and the connection with the complex dynamical state of this cluster is not trivial, and it is beyond the scope of this work. Here we just mention that this result is in agreement to what is observed in general for radio mini-halos \citep[][]{Ignesti_2020}.
\begin{table}[h!]
\centering
\begin{tabular}{lc}
\hline
 SMptp: fitting method&$k$\\
\hline
 Least squares&0.66$\pm$0.19\\
 BCES orthogonal&0.85$\pm$0.14\\
 BCES bisector& 0.88$\pm$0.12\\
 LinMix&0.75$\pm$0.19\\
 \hline
 MCptp: number of iterations&$k$\\
 \hline
 50 &1.09$\pm$0.28\\
 100& 1.03$\pm$0.28\\
200& 1.07$\pm$0.26\\
500& 1.06$\pm$0.26\\
 \hline
\end{tabular}
\caption{\label{res} Results of the different ptp analysis.}
\end{table}
\section{Conclusions}
We introduced PT-REX, a software to estimate the point-to-point trend between radio and X-ray surface brightness in extended sources. We presented here a set of tools to perform this analysis and we discussed how to run them effectively for a variety of scientific problems. We also introduced the Monte Carlo ptp analysis, which allows to explore the parameter space of the scaling by combining numerous random sampling of the source. This method is advised when studying small or poorly-resolved sources. Finally, we showed how to use PT-REX by studying the $I_\text{R}$-$I_\text{X}$ trend in the cluster RXJ1347.5-1145. Despite the low resolution of the radio image, we could reliably estimate a linear scaling between radio and X-ray emission ($k=1.06\pm0.26$) that indicates a strong bond between the thermal gas and the radio-emitting plasma. With PT-REX similar studies can be easily conducted on a variety of science cases to derive some useful (and sometimes unexpected) insights into their physics.
\section*{Acknowledgments}
AI thanks G. Brunetti and M. Sereno for the useful discussions, M. Gitti for providing the VLA image of RXJ1347.5-1145, and L. Bruno for having tested the code with infinite patience. This research made use of Astropy\footnote{{\tt http://www.astropy.org}}, a community-developed core Python package for Astronomy \citep[][]{Astropy_2013, Astropy_2018}, SciPy \citep[][]{Scipy_2020} and APLpy, an open-source plotting package for Python \citep[][]{Robitaille_2012}.
\bibliographystyle{elsarticle-num-names}
\bibliography{bibliography}
\end{document}